\input epsf
\documentstyle[preprint,aps]{revtex}     
\begin{document}
\draft
\title{X-Ray Scattering at FeCo(001) Surfaces and the\\
Crossover between Ordinary and
Normal Transitions}
\author{Uwe Ritschel\thanks{e-mail: uwe@theo-phys.uni-essen.de}}
\address{Fachbereich Physik, Universit\"at GH Essen, 45117 Essen
(Germany) and\\
Fachbereich Physik, Carl-von-Ossietzky-Universit\"at Oldenburg,
26111 Oldenburg (Germany)}
\maketitle
\narrowtext
\begin{abstract}
In a recent experiment by Krimmel et al. [Phys. Rev. Lett. {\bf 78}, 3880 (1997)], the 
critical behavior of FeCo near a (001) surface was studied
by x-ray scattering. 
Here the experimental data are reanalyzed, taking into account
recent {\it theoretical}
results on order-parameter profiles in the
crossover regime between ordinary and normal transitions.
Excellent agreement between theoretical expectations
and the experimental results is found.
\end{abstract}
Through the years, binary alloys like Fe$_3$Al or
FeCo  have provided one of the main test objects
for theories of surface critical phenomena \cite{Binder}. 
Some time ago, the experiment
of Mail\"ander et al. \cite{Mailander,Dosch} very impressively verified earlier
theoretical predictions of Dietrich and Wagner \cite{Dietrich} on the
influence of the order parameter and the correlation function
on scattering intensities near criticality. 
More recently, Krimmel et al. \cite{Krimmel}
studied the surface critical behavior of FeCo near a (001) surface. 
FeCo at or close to the 
ideal stoichiometry (50$\%$ Fe and 50$\%$ Co)
undergoes a continuous disorder-order transition at about
$T_c=900 - 1000$\,K (the critical
temperature depends on the precise stoichiometry).
In the high-temperature disordered phase the two species
are distributed randomly on the sites of a body-centered cubic lattice
(A2 phase). Below $T_c$, Fe and Co segregate on
the two sublattices (B2 phase) \cite{Collins}.
The order parameter is proportional to the difference between
the sublattice concentrations. As a consequence, the system
can be modelled by an Ising antiferromagnet \cite{Binder86},
and the corresponding bulk universality class 
of the transition is the one of the (ferromagnetic)
Ising model. Earlier experiments on the
bulk critical behavior provided good
evidence for this scenario \cite{Oyedele}. 

Concerning the surface critical behavior, due to missing neighbors
surface spins have a reduced tendency to order such that
these systems should belong to the surface universality class
of the ``ordinary transition'' \cite{Binder}. 
Like the previous work on Fe$_3$Al \cite{Mailander},
the work of Krimmel et al. \cite{Krimmel} again provided 
good evidence for this. The temperature dependence of the order parameter 
near the (001) surface, measured by surface-sensitive
evanescent wave scattering \cite{Dosch}, can be described by a power law
with the exponent $\beta_1\simeq 0.79\pm 0.1$ \cite{Krimmel}, 
a value that agrees well with theoretical expectations \cite{Binder,Comment1}.
However, both experiments \cite{Krimmel,Mailander,Dosch2}
also revealed the existence of some residual
long-range order near the surface for $\tau\equiv (T-T_c)/T_c >0$,
which, at the first glance, seemed
incompatible with the scenario of the ordinary transition.

Theoretically the A2-B2 transition
like the one in FeCo was studied in the framework of a
lattice spin model \cite{Binder86} by Schmid \cite{Schmid}. 
The important 
result of Schmid was that in a system with {\it non-ideal stoichiometry}
segregation effects near a (001) surface 
generate an ordering (staggered) surface field $h_1$ that, even for temperatures
$\tau > 0$, stabilize a residual order $m_1$ 
in the surface layer, in consistency with the observations
of Refs.\,\cite{Krimmel,Mailander,Dosch2}.

Taking into account the theoretical and experimental findings discussed
above, one has to conclude that experiments of the kind reported
in Ref.\,\cite{Krimmel} are generically carried out in the crossover
regime between ``ordinary'' ($h_1=0$) and ``normal'' ($h_1=\infty$)
transitions, in which the order parameter, in general, is a function $m(z)$
of the distance $z$ from the surface.
At the ordinary transition,
$m(z)$ vanishes identically for $\tau>0$.
For $\tau<0$ the surface orders passively as $m_1\sim |\tau|^{\beta_1}$
and the crossover from surface to bulk behavior,
$m_{\rm bulk}\sim |\tau|^{\beta}$, is described
by a profile of the form
$m(z)\sim z^{(\beta_1-\beta)/\nu}$ \cite{Gompper}.    
At the normal transition, on the other hand,
the surface is completely ordered by a 
strong surface field. With increasing $z$, $m(z)$ monotonically
decays towards the bulk value.
This decay is described
by the power-law $\sim z^{-\beta/\nu}$ for $a\ll z\ll \xi$
and by $\sim \exp(-z/\xi)$ for $z\gg \xi$ (where the critical
exponents have their standard meaning, $a$ represents some
microscopic length (lattice constant), and $\xi\sim |\tau|^{-\nu}$
is the bulk correlation length). 

As discussed recently \cite{Ritschel}, any finite $h_1$ leads to a
{\it non-monotonic} profile $m(z)$. The surface field provides an additional
length scale $l_1\sim h_1^{-\nu/\Delta_1}$. At bulk criticality
and for $a\ll z< l_1$, $m(z)$ increases as $\sim
h_1\,z^{(\Delta_1-\beta)/\nu}$,
before it decays to zero as $\sim z^{-\beta/\nu}$ for $z>l_1$.
Slightly away
from $T_c$, the scenario holds for $z< \xi$, and farther away
from the surface an exponential
decay sets in. For instance for $\tau>0$ and $\xi<l_1$, the 
case which turns out to be relevant for the experiment,
$m(z)$ increases up to $z\simeq \xi$ and then 
crosses over to an exponential decay\cite{Czerner}.

Taking this crossover scenario at face value,
what are the consequences for
the measurable quantities in x-ray scattering experiments?
A nonvanishing order parameter near the surface makes itself
felt in form of an additional
superstructure peak in the (001)
direction of the reciprocal lattice \cite{Krimmel,Comment3,Reichert}.
The data of Ref.\,\cite{Krimmel} 
for the integrated peak intensity $\widehat I$
are displayed in Fig.\,1 (symbols with
error bars).

In order to calculate $\widehat I$ from the profile $m(z)$,
one may to a first approximation assume that the 
intensity distribution in the peak is given by \cite{Krimmel}
\begin{equation}
I(Q_z)=\left| \int_0^{\infty} dz\, m(z)\, e^{i\,Q_zz}\right|^2\>.
\end{equation}
Then for the integrated intensity the result reads
\begin{equation}\label{hati}
\widehat I  =\int_{-\infty}^{\infty}\,dQ_z\>I(Q_z)=\int_0^{\infty}\,dz\,
\left[m(z,h_1,\tau)\right]^2\,.
\end{equation}

For the profile $m(z)$, I choose the simple ansatz
\begin{equation}\label{ansatz}
m(z,h_1,\tau)\propto 
\left\{\begin{array}{ll}
h_1\,z^{(\Delta_1-\beta)/\nu}, & z<\xi\\
h_1\,\xi^{(\Delta_1-\beta)/\nu}\,e^{-z/\xi+1},&z>\xi
\end{array}\right.
\end{equation} 
which is sufficient for my present purpose.
It yields the $\tau$ dependence of the integral
(\ref{hati}) for weak $h_1$ in the regime
$l_1 \gg \xi$.
This is the situation encountered in the experiment Ref.\,\cite{Krimmel},
because only a weak $h_1$ is consistent with the observation
of ``ordinary'' behavior for $\tau <0$ \cite{Ritschel}.
Inserting (\ref{ansatz}) in (\ref{hati}),
the $\tau$ dependence of $\widehat I$ turns out as
\begin{equation}\label{power}
\widehat I \sim \tau^{2(\beta-\Delta_1)-\nu}\,.
\end{equation}
With the literature values $\beta=0.32$, $\nu=0.63$, and
$\Delta_1=0.46$ \cite{Binder}, this  
means $\widehat I\sim \tau^{-0.91}$,
in excellent agreement with the experimental data as demonstrated
by the solid line in Fig.\,1.
With a similar ansatz it is straightforward to show that
in the case $\xi \gg l_1$, i.e., in a regime much closer to $T_c$
and probably not resolved in Ref.\,\cite{Krimmel},
the $\tau$ dependence
of the intensity is described by $\widehat I \sim
\tau^{2\beta-\nu}$ \cite{Comment5}.  

In contrast, 
Krimmel et al. \cite{Krimmel} assumed an exponential decay
\begin{equation}\label{exponential}
m(z)\sim m_1\,\exp(-z/\xi)
\end{equation}
of the profile and explained the vanishing of residual order
above $T_c$ with a relatively strong dependence of $m_1$ on the
temperature.
The $\tau$ dependence of $m_1$ is described by
\begin{equation}\label{taudep}
m_1\sim A-B\,\tau^{-\gamma_{11}}\>,
\end{equation}
with the literature value
$\gamma_{11}\simeq - 0.33$ \cite{Binder}
and where
the constants $A$ and $B$ were used as fit parameters in Ref.\,\cite{Krimmel}.
In my approach, on the other hand, the decay 
according to (\ref{taudep}) is assumed to be slow
and negligible for the temperature range probed in the experiment.

Just from the data it is not possible
to discriminate between the two fits and, thus, the underlying
theoretical models. However, there are
(at least) two reasons why the scenario proposed in this
work should be correct:
\begin{itemize}
\item The simulation results of Schmid \cite{Schmid} indicate a rather
slow variation of $m_1$ as a function of $\tau$.
More precisely, the data shown e.g. in Fig.\,7.\,c of Ref.\,\cite{Schmid}
reveal that for $\tau\simeq 0.3$, $m_1$ is still
at about half of its value at $\tau= 0$. In the experiment, on the other hand, 
the signal vanishes for $\tau\gtrsim 0.01$. Hence, it is very
unlikely that the decay of long-range order
above $T_c$ is attributable to the
decay of $m_1$.
\item From experiments on similar systems \cite{Guttman} it is
known that the $\tau$-dependence of the measured correlation length
agrees extremely well with the theoretical
expectation $\xi\sim |\tau|^{-0.63}.$
With the exponential fit, where
the correlation length enters as a fit parameter for the $Q_z$ dependence
of the intensity profiles, one
obtains an increasing $\xi$ upon approaching $T_c$ (see Fig.\,4.b in Ref.\,\cite{Krimmel}),
but with a $\tau$-dependence that is not consistent with the theory.
In the approach proposed in this work the correct theoretical
$\tau$-dependence enters directly in the ansatz (\ref{ansatz})
for the profile, and, as the solid line in Fig.\,1 shows,
the data in this case are consistent with the theory.
\end{itemize}

Finally I should like to mention that a power law similar to
(\ref{power}) was derived in the context of light scattering
experiments by Franck \cite{Franck}.
In this case the reflectivity of binary mixtures near the consolute
point is given the {\it first moment} of
$m(z)$ and behaves as $\sim \tau^{\beta-2\nu-\Delta_1}$.
In Ref.\,\cite{Franck} this result was derived from scaling considerations,
without explicit knowledge of the underlying order-parameter profiles.

To conclude, I have presented evidence that the results of the
experiment of Krimmel et al. \cite{Krimmel} can be explained with
the crossover profiles introduced in Ref.\,\cite{Ritschel}.
A detailed fit of the fine structure of the superstructure
reflections
as presented in Fig.\,3 of Ref.\,\cite{Krimmel}
remains to be carried out with the crossover profiles.\\
{\small {\it Acknowledgements}:
I should like to thank W. Donner, H. Dosch, and S. Krimmel
for explaining me their experiment in detail as well as
for leaving me their original data.}

\newpage
\noindent
{\Large \bf Figure Caption} 
\\[5mm]
{\small {\bf Fig.\,1:} 
The integrated scattering intensity $\widehat I$ above the
bulk critical point. The inset shows the same data in double-logarithmic
form. The experimental data of Krimmel et al. \cite{Krimmel}
are represented by the full circles. (A background intensity
of 4000 was subtracted.)
The fit proposed in this
Communication is represented by the full line, the original
fit of Ref.\,[1] by the dashed line.}
\newpage
\vspace*{5cm}
\hspace*{1.5cm}\def\epsfsize#1#2{0.9#1}
\epsfbox{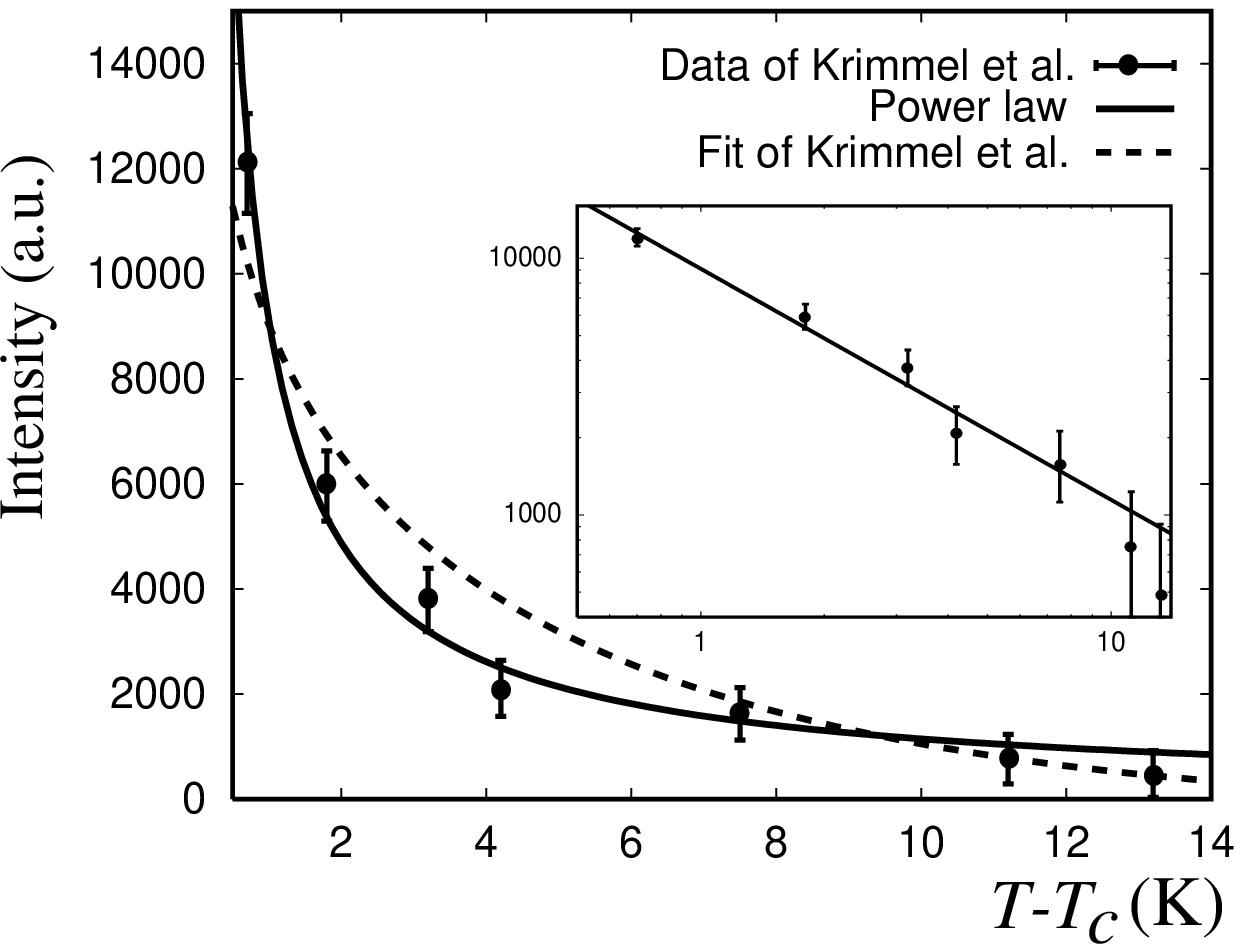}\\[1cm]
{\large \bf} Fig.1


\begin{thebibliography} {99}
\bibitem{Comment1} The bulk universality class of the Ising model
splits up in several surface universality classes, corresponding
to different fixed points of the renormalization-group flow \cite{Binder}.
In the present work, however, only the exponents of the ordinary transition
are relevant, and, hence, additional indices at the exponents are omitted.
\bibitem{Comment3} Due to lattice distortions
near the surface, this peak has a ``fine structure'' in form
of two clearly distinguishable subpeaks \cite{Krimmel} .
However, this fine structure should not affect much the integrated
intensity $\widehat I$ discussed in the text.
\bibitem{Comment5} Since in the regime $l_1\ll \xi$ the exponent
that governs $\widehat I$ is $2\beta-\nu \simeq 0$, one may speculate
that the clustering of data around $T_c$ observed in
Fig.\,4.b of Ref.\,\cite{Krimmel} could be due to the 
very weak or vanishing $\tau$ dependence
in this limit. However, the error bars on the data are
too large to yield solid support for this hypothesis.
\end{thebibliography}
\end{document}